\documentclass[aps,prd,superscriptaddress,floatfix,showpacs]{revtex4}

\usepackage{verbatim}

\usepackage{amsmath}
\usepackage{amsfonts}
\usepackage{amssymb}
\usepackage{graphicx}
\usepackage{bm}
\usepackage{color}
\usepackage{epsf}

\usepackage{psfrag}

\def\bea{\begin{eqnarray}}
\def\eea{\end{eqnarray}}
\def\beas{\begin{eqnarray*}}
\def\eeas{\end{eqnarray*}}
\def\beqas{\begin{eqnarray*}}
\def\eqas{\end{eqnarray*}}
\def\beq{\begin{equation}}
\def\eeq{\end{equation}}
\def\beqd{\begin{displaymath}}
\def\eeqd{\end{displaymath}}
\def\eqd{\end{displaymath}}

\def\slashchar#1{\setbox0=\hbox{$#1$}
   \dimen0=\wd0
   \setbox1=\hbox{/} \dimen1=\wd1
   \ifdim\dimen0>\dimen1
      \rlap{\hbox to \dimen0{\hfil/\hfil}}
      #1
   \else\begin{eqnarray}
      \rlap{\hbox to \dimen1{\hfil$#1$\hfil}}
      /
   \fi}

\begin{document}
\title
{Hard photoproduction of a diphoton with a large invariant mass}
\author{A.~Pedrak}
\affiliation{ National Center for Nuclear Research (NCBJ), 00681 Warsaw, Poland}
\author{ B.~Pire}
\affiliation{ Centre de Physique Th\'eorique, \'Ecole Polytechnique,
CNRS,  91128 Palaiseau,     France  }
\author{ L.~Szymanowski}
\affiliation{ National Center for Nuclear Research (NCBJ), 00681 Warsaw, Poland}
\author{ J.~Wagner}
\affiliation{  National Center for Nuclear Research (NCBJ), 00681 Warsaw, Poland}
\begin{abstract}
 The electromagnetic probe has proven to be a very efficient way to access the $3-$dimensional structure of the nucleon, particularly thanks to the exclusive Compton processes. We explore the hard photoproduction of a large invariant mass diphoton in the kinematical regime where the diphoton is nearly forward and its invariant mass is the hard scale enabling to factorize the scattering amplitude in terms of generalized parton distributions. This amplitude has a very peculiar and interesting analytic structure. We calculate unpolarized cross sections and the angular asymmetry triggered by a linearly polarized photon beam.   
\end{abstract}
\maketitle
\section{Introduction}
The last twenty years have witnessed a tremendous progress in the understanding of hard exclusive scattering in the framework of the QCD collinear factorization  of hard amplitudes in specific kinematics in terms of generalized parton distributions (GPDs) and hard perturbatively calculable coefficient functions \cite{historyofDVCS, gpdrev}.

In this paper, we study the exclusive photoproduction of two photons on a   unpolarized proton or neutron target
\begin{equation}
\gamma(q,\epsilon) + N(p_1,s_1) \rightarrow \gamma(k_1,\epsilon_1) +  \gamma(k_2,\epsilon_2)+ N'(p_2,s_2)\,,
\label{process}
\end{equation}
 in the kinematical regime of large invariant diphoton mass  $M_{\gamma\gamma}$ of the final photon pair and small momentum transfer $t =(p_2-p_1)^2$ between the initial and the final nucleons. Roughly speaking, these kinematics means a moderate to large, and approximately opposite, transverse momentum of each  final photon. This reaction has a number of interesting features. First, it is a purely electromagnetic process at Born order - as are deep inelastic scattering (DIS), deeply virtual Compton scattering (DVCS) and timelike Compton scattering (TCS) - and, although there is no deep understanding of this fact, this property is usually accompanied by early scaling. Second, the process is insensitive to gluon GPDs because of the charge symmetry of the two photon final state. This may help to reduce QCD next to leading order corrections (which we do not calculate here) since they are often more important for gluon initiated partonic processes than for quark initiated ones. Third, there is also no contribution from the badly  known  chiral-odd quark distributions. This study enlarges the range of $2 \to 3$ reactions analyzed in the framework of collinear QCD factorization \cite{2to3, elb}.  
\section{Kinematics}
\label{Sec:Kinematics}
Let us first present the kinematics of the process (\ref{process}). 
We decompose every momenta on a Sudakov basis  as
\begin{equation}
\label{sudakov1}
v^\mu = a \, n^\mu + b \, p^\mu + v_\bot^\mu \,,
\end{equation}
with $p$ and $n$ the light-cone vectors
\begin{equation}
\label{sudakov2}
p^\mu = \frac{\sqrt{s}}{2}(1,0,0,1)\,,\qquad n^\mu = \frac{\sqrt{s}}{2}(1,0,0,-1)\,, \qquad p\cdot n = \frac{s}{2}\,,
\end{equation}
and
\begin{equation}
\label{sudakov3}
 v^+ = v.n  \,, \qquad v_\bot^\mu = (0,v^x,v^y,0) \,, \qquad v_\bot^2 = -\vec{v}_t^2\,.
\end{equation}

The particle momenta read
\begin{eqnarray}\label{impfinc}
 p_1^\mu &=& (1+\xi)\,p^\mu + \frac{M^2}{s(1+\xi)}\,n^\mu~, \quad p_2^\mu = (1-\xi)\,p^\mu + \frac{M^2+\vec{\Delta}^2_t}{s(1-\xi)}n^\mu + \Delta^\mu_\bot\,, \quad q^\mu = n^\mu ~,\nonumber \\
k_1^\mu &=& \alpha_1 \, n^\mu + \frac{(\vec{p}_t-\vec\Delta_t/2)^2}{\alpha_1 s}\,p^\mu + p_\bot^\mu -\frac{\Delta^\mu_\bot}{2}~,~~
 k_2^\mu = \alpha_2 \, n^\mu + \frac{(\vec{p}_t+\vec\Delta_t/2)^2}{\alpha_2 s}\,p^\mu - p_\bot^\mu-\frac{\Delta^\mu_\bot}{2}\,,
\end{eqnarray}
where 
$M$ is the mass of the nucleon and $\xi$ is the skewness parameter.  We define $\Delta^\mu = p_2^\mu-p_1^\mu$ and $P^\mu = \frac{1}{2} (p_1^\mu+p_2^\mu)$.

The total center-of-mass energy squared of the $\gamma$-N system is
\begin{equation}
\label{energysquared}
S_{\gamma N} = (q + p_1)^2 = (1+\xi)s + M^2\,.
\end{equation}
The  Mandelstam invariants read
\begin{eqnarray}
\label{s't'u'}
t &=& (p_2 - p_1)^2 = -\frac{1+\xi}{1-\xi}\vec{\Delta}_t^2 -\frac{4\xi^2M^2}{1-\xi^2}\,, \\
M_{\gamma\gamma}^2 &=& ~(k_1 +k_2)^2 = ~ 2 \xi \, s \left(1 - \frac{ 2 \, \xi \, M^2}{s (1-\xi^2)}  \right) - \vec{\Delta}_t^2 \frac{1+\xi}{1-\xi}\,, \\
\label{t'}
- t'&=& -(k -q)^2 =~\frac{(\vec p_t-\vec\Delta_t/2)^2}{\alpha_1} \;,\\
\label{u'}
- u'&=&- (k_2-q)^2= ~\frac{(\vec p_t+\vec\Delta_t/2)^2}{\alpha_2}
 \; .
\end{eqnarray}
The simplified kinematical relations used to calculate the hard coefficient functions are :
\begin{equation}
\label{simplkin}
\alpha_1 + \alpha_2 = 1~~~~,~~~~\alpha_1 \alpha_2 = \frac{\vec{p_t}^2}{M^2_{\gamma\gamma}}~~~~,~~~~\xi = \frac{\tau}{2-\tau} ~~~~,~~~~\tau = \frac{M^2_{\gamma\gamma}-t}{S_{\gamma N}-M^2}\,.
\end{equation}

\section{The scattering amplitude}
\begin{figure}
\includegraphics[width=8cm]{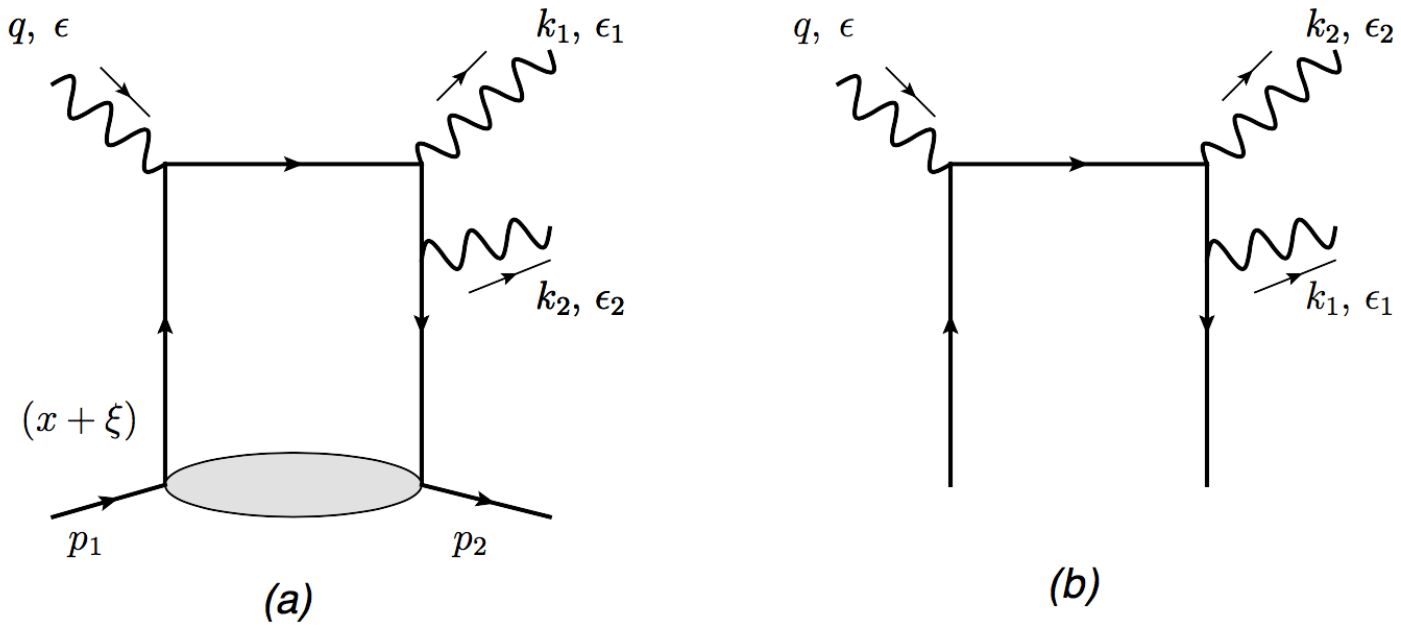}
\includegraphics[width=8cm]{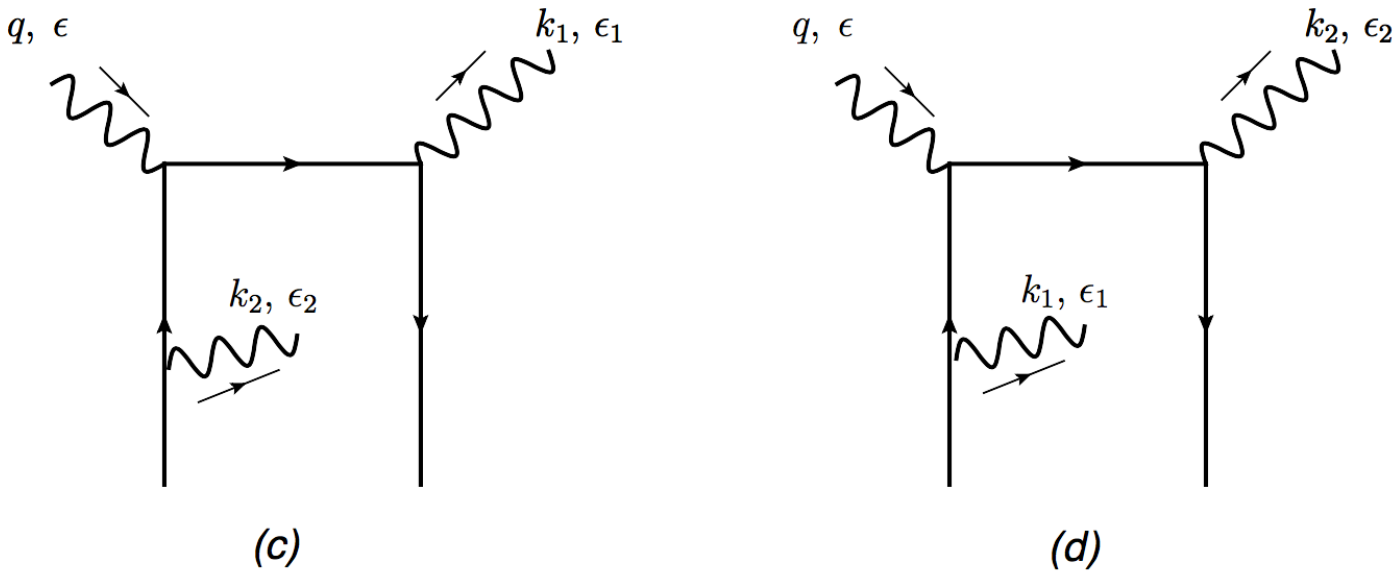}

\vspace{-2cm}\includegraphics[width=8cm]{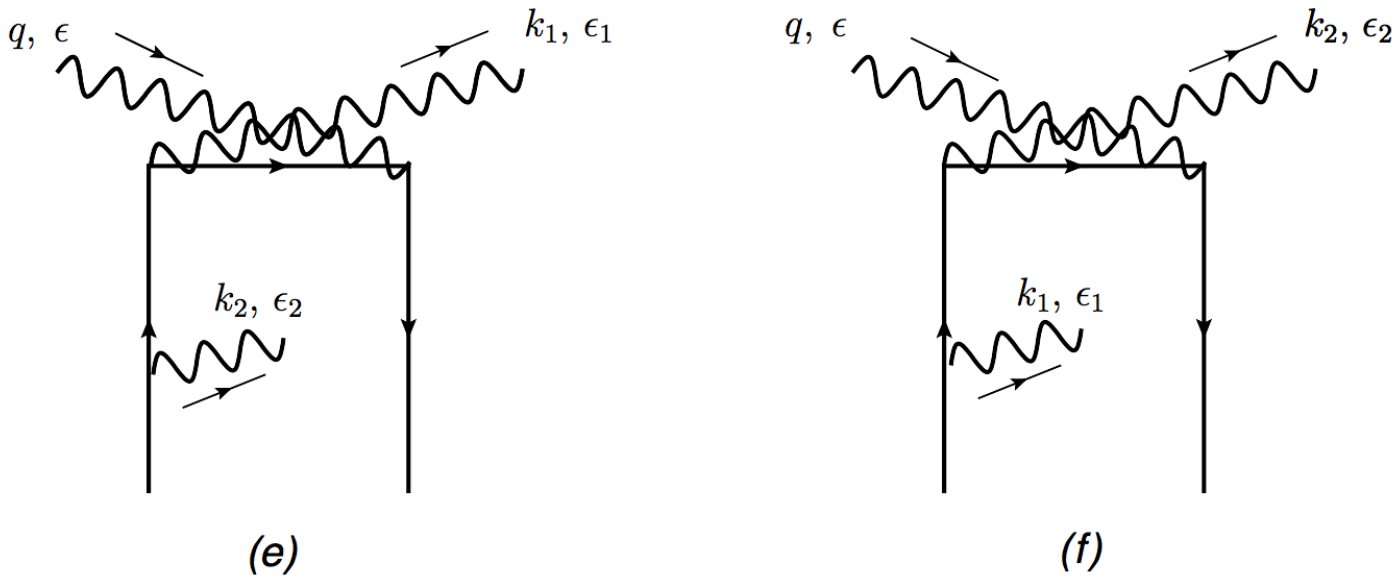}
\vspace{-1cm}
\caption{Feynman diagrams contributing to the coefficient functions of the process $\gamma N \to \gamma \gamma N'$}
\label{feyndiageu3g}
\end{figure}

Factorization allows to write the scattering amplitude as
\begin{eqnarray}
&\mathcal{T} & =
\frac{1}{4}\int_{-1}^1 dx \sum_q\int dz^- \exp(ixz^-P^+)
\left\langle p_2, s_2\left|
\overline{\psi}_q\left(-\frac{z}{2}\right)
\left[
CF^V_q\not{n}+CF^A_q\not{n}\gamma^5
\right]
\psi_q\left(\frac{z}{2}\right)
\right|p_1, s_1\right\rangle\\
&&=\frac{1}{2}\;\frac{1}{2P^+}
\int_{-1}^1 dx\sum_q\left[
CF^V_q(x,\xi)\left(
 H^q(x,\xi)\bar{U}(p_2, s_2)\not{n} U(p_1,s_1)+
 E^q(x,\xi)\bar{U}(p_2, s_2)\frac{i\sigma^{\mu\nu}\Delta_\nu n_\mu}{2M}U(p_1, s_1)
\right)+
\right.\nonumber \\
&&\left. ~~~~~~~~~~~~~~~~~~~~~
+CF^A_q(x,\xi)\left(
\tilde{H}^q(x,\xi)\bar{U}(p_2, s_2)\not{n}\gamma^5 U(p_1, s_1)+
\tilde{E}^q(x,\xi)\bar{U}(p_2, s_2)\frac{i\gamma_5(\Delta\cdot n)}{2M}U(p_1, s_1)
\right)
\right]\, ,
\end{eqnarray}
with the coefficient functions calculated from the diagrams of Fig. 1 \footnote{Symmetries help to reduce the number of diagrams to be calculated since
diagram b = diagram a ( $k_1,\epsilon_1 \leftrightarrow k_2, \epsilon_2)$;
diagram f = diagram e ( $k_1,\epsilon_1 \leftrightarrow k_2, \epsilon_2)$;
diagram c = diagram d ( $k_1,\epsilon_1 \leftrightarrow k_2, \epsilon_2)$;
diagram e = diagram a ( $x \leftrightarrow -x)$. One thus only needs to calculate diagram (a) and (d).
} 
as
\begin{eqnarray}
iCF^V_q&=& Tr[i\mathcal{M}\not p]=
-i e_q^3\left[A^V\left(\frac{1}{D_1(x)D_2(x)}+\frac{1}{D_1(-x)D_2(-x)}\right)+\right.\\
&&B^V\left(\frac{1}{D_1(x)D_3(x)}+\frac{1}{D_1(-x)D_3(-x)}\right)+
\left.C^V\left(\frac{1}{D_2(x)D_3(-x)}+\frac{1}{D_2(-x)D_3(x)}\right)\right] \,,
\nonumber\\
iCF^A_q&=& Tr[i\mathcal{M}\gamma^5\not p]=
-i e_q^3\left[A^A\left(\frac{1}{D_1(x)D_2(x)}-\frac{1}{D_1(-x)D_2(-x)}\right)+\right.
\left.
B^A\left(\frac{1}{D_1(x)D_3(x)}-\frac{1}{D_1(-x)D_3(-x)}\right)
\right]
\nonumber
\end{eqnarray}
where ${\mathcal M}\not p$ and ${\mathcal M}\gamma^5\not p$ are contributions of the hard part of scattering amplitude projected on vector and axial vector   Fierz structures, 
 $e_q=Q_q |e|$  and the denominators read
 \begin{eqnarray}
&&D_1(x)=s(x+\xi+i\varepsilon) ~~~,~~~
D_2(x)=s\alpha_2(x-\xi+i\varepsilon)~~~,~~~
D_3(x)=s\alpha_1(x-\xi+i\varepsilon)\,.
\end{eqnarray}
The tensorial structure can be written for the vector part as
\begin{eqnarray}
A^V=2s(V_{k_1}-V_p+\frac{1+\alpha_2}{\alpha_1}V_{k2})~,~
B^V=2s(-V_{k2}+V_p-\frac{1+\alpha_1}{\alpha_2}V_{k_1})~,~
C^V=2s((\alpha_2-\alpha_1)V_p+V_{k_2}-V_{k1})\,,
\end{eqnarray}
with 
\begin{eqnarray}
V_{k_1}=(\epsilon_\bot(q)\cdot\epsilon^*_\bot(k_1))
(p_\bot\cdot\epsilon^*_\bot(k_2))~,~
V_{k_2}=(\epsilon_\bot(q)\cdot\epsilon^*_\bot(k_2))
(p_\bot\cdot\epsilon^*_\bot(k_1))~,~
V_{p}=(\epsilon^*_\bot(k_1)\cdot\epsilon^*_\bot(k_2))
(p_\bot\cdot\epsilon_\bot(q))\,,
\end{eqnarray}
while the axial part reads
\begin{eqnarray}
A^A=4i\left(A_{k_1}+\frac{1+\alpha_2}{\alpha_1}A_{k_2}-A_p\right)~,~ B^A=4i\left(-\frac{1+\alpha_1}{\alpha_2}A_{k_1}-A_{k_2}+A_p\right) \,,
\end{eqnarray}
with
\begin{eqnarray}
A_{k_1}=p_\bot\cdot\epsilon^*_\bot(k_2)\epsilon^{pn\epsilon_\bot(q)\epsilon_\bot^*(k_1)}~,~
A_{k_2}=p_\bot\cdot\epsilon^*_\bot(k_1)\epsilon^{pn\epsilon_\bot(q)\epsilon_\bot^*(k_2)}~,~
A_{p}=\epsilon^*_\bot(k_1)\cdot\epsilon^*_\bot(k_2)\epsilon^{pn\epsilon_\bot(q)p_\bot}\,.
\end{eqnarray}

The scattering amplitude is written in terms of generalized Compton form factors $\mathcal{H}^q(\xi)$, 
$\mathcal{E}^q(\xi)$, $\tilde {\mathcal{H}}^q(\xi)$ and 
$\tilde {\mathcal{E}}^q(\xi)$ 
as
\begin{eqnarray}
&\mathcal{T}=&
\frac{1}{2s}
\sum_q
\left[
\left(
\mathcal{H}^q(\xi)\bar{U}(p_2)\not{n} U(p_1)+
 \mathcal{E}^q(\xi)\bar{U}(p_2)\frac{i\sigma^{\mu\nu}\Delta_\nu n_\mu}{2M}U(p_1)
\right)+\right. \nonumber\\
&&
\left.
\left(
\tilde{\mathcal{H}}^q(\xi)\bar{U}(p_2)\not{n}\gamma^5 U(p_1)+
\tilde{\mathcal{E}}^q(\xi)\bar{U}(p_2)\frac{i\gamma_5(\Delta\cdot n)}{2M}U(p_1)
\right)
\right]\,,
\end{eqnarray}

where

\begin{eqnarray}
\mathcal{H}^q(\xi)&=&\int^1_{-1}dx CF^V_q(x,\xi)H^q(x,\xi)=
(-e_q^3)\left[A^V\mathcal{H}^q_{A^V}(\xi)+B^V\mathcal{H}^q_{B^V}(\xi)+C^V\mathcal{H}^q_{C^V}(\xi)\right]\, ,\nonumber
\\
& = &(-e_q^3) (\alpha_1A^V+\alpha_2B^V) \frac{i\pi}{\xi s^2\alpha_1\alpha_2}(H^q(\xi,\xi)+H^q(-\xi,\xi))\, ,
\\
\mathcal{E}^q(\xi)&=&\int^1_{-1}dx CF^V_q(x,\xi)E^q(x,\xi)=
(-e_q^3)\left[A^V\mathcal{E}^q_{A^V}(\xi)+B^V\mathcal{E}^q_{B^V}(\xi)+C^V\mathcal{E}^q_{C^V}(\xi)\right] \, ,\nonumber
\\
& = &(-e_q^3) (\alpha_1A^V+\alpha_2B^V) \frac{i\pi}{\xi s^2\alpha_1\alpha_2}(E^q(\xi,\xi)+E^q(-\xi,\xi))\, ,
\\
\tilde{\mathcal{H}}^q(\xi)&=&\int^1_{-1}dx CF^A_q(x,\xi)\tilde{H}^q(x,\xi)=
(-e_q^3)\left[A^A\tilde{\mathcal{H}}^q_{A^A}(\xi)+B^A\tilde{\mathcal{H}}^q_{B^A}(\xi)+C^A\tilde{\mathcal{H}}^q_{C^A}(\xi)\right] \, ,
\\
\tilde{\mathcal{E}}^q(\xi)&=&\int^1_{-1}dx CF^A_q(x,\xi)\tilde{E}^q(x,\xi)=
(-e_q^3)\left[A^A\tilde{\mathcal{E}}^q_{A^A}(\xi)+B^A\tilde{\mathcal{E}}^q_{B^A}(\xi)+C^A\tilde{\mathcal{E}}^q_{C^A}(\xi)\right] \, ,
\end{eqnarray}
with
\begin{eqnarray}
&&\tilde{\mathcal{H}}^q_{A^A}(\xi)=-\frac{i\pi}{\xi\alpha_2s^2}(\tilde{H}^q(\xi,\xi)-\tilde{H}^q(-\xi,\xi)) \, , \\
&&\tilde{\mathcal{H}}^q_{B^A}(\xi)=-\frac{i\pi}{\xi\alpha_1s^2}(\tilde{H}^q(\xi,\xi)-\tilde{H}^q(-\xi,\xi)) \, , \\
&&\tilde{\mathcal{H}}^q_{C^A}(\xi)=0\, ,\\
&&\tilde{\mathcal{E}}^q_{A^A}(\xi)=-\frac{i\pi}{\xi\alpha_2s^2}(\tilde{E}^q(\xi,\xi)-\tilde{E}^q(-\xi,\xi)) \, , \\
&&\tilde{\mathcal{E}}^q_{B^A}(\xi)=-\frac{i\pi}{\xi\alpha_1s^2}(\tilde{E}^q(\xi,\xi)-\tilde{E}^q(-\xi,\xi)) \, , \\
&&\tilde{\mathcal{E}}^q_{C^A}(\xi)=0\, .
\end{eqnarray}
It is a remarkable result that the leading twist leading order amplitude is proportional to valence quark  generalized parton distributions taken at the border value $x=\pm \xi$. 

Summing (averaging) over the final (initial) nucleon spin ($s_1,s_2$) yields the squared scattering amplitude at $\Delta_T=0$ as
\begin{eqnarray}
&
\frac{1}{2}\sum_{s_1,s_2}|\mathcal{T}|^2=
&
\frac{1}{4}
\left[
 (1-\xi^2)
\sum_q\mathcal{H}^q \sum_q\mathcal{H}^{q*}+
(-\xi^2)(\sum_q\mathcal{H}^q \sum_q\mathcal{E}^{q*}+\sum_q\mathcal{E}^q \sum_q\mathcal{H}^{q*})+
\left(
\frac{\xi^4}{1-\xi^2}
\right)\sum_q\mathcal{E}^q \sum_q\mathcal{E}^{q*}
\right.\nonumber
\label{tau2}
\\
&&
\left.
+(1-\xi^2)
\sum_q\tilde{\mathcal{H}^q} \sum_q\tilde{\mathcal{H}}^{q*}+
(-\xi^2)
(\sum_q\tilde{\mathcal{H}}^q \sum_q\tilde{\mathcal{E}}^{q*}+
\sum_q \tilde{\mathcal{E}}^q \sum_q\tilde{\mathcal{H}}^{q*}
)+
\left(
\frac{\xi^4}{1-\xi^2}
\right)
\sum_q \tilde{\mathcal{E}}^q \sum_q\tilde{\mathcal{E}}^{q*}
\right]\,.
\end{eqnarray}
 \begin{figure}[h!]
\centering
\includegraphics[height=7cm]{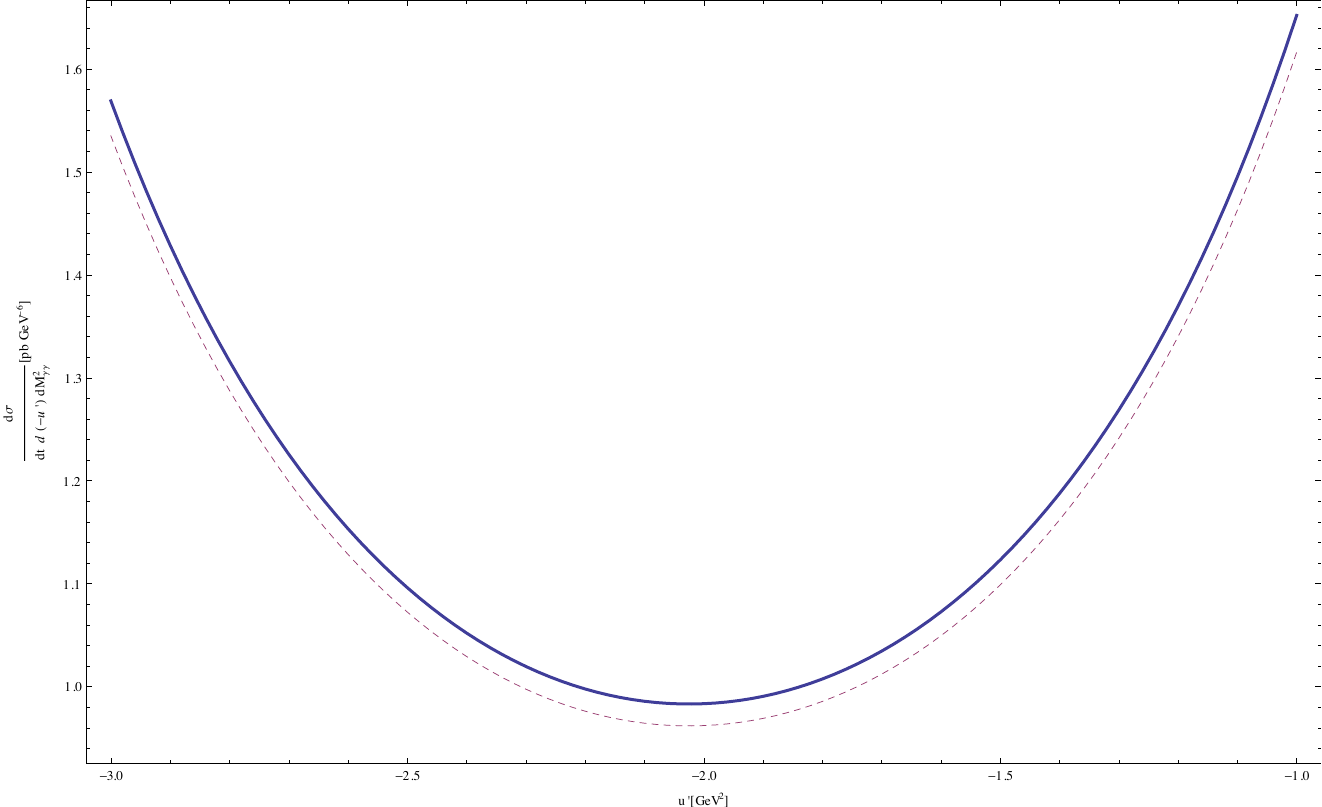}
\caption{The $u'$ dependence of the unpolarized differential cross section $\frac{d\sigma}{dM_{\gamma\gamma}^2du'dt} $ at $t=t_{min}$ and $S_{\gamma N} = 20$ GeV$^2$, for $M_{\gamma\gamma}^2= 4$ GeV$^2$ with the quark GPDs modeled as in \cite{GK} (solid curve)  and in \cite{VGG}  (dashed curve) for a proton target. }
\label{VGG_vs_GK}
\end{figure}
 \begin{figure}[h!]
\centering
\includegraphics[height=4.5cm]{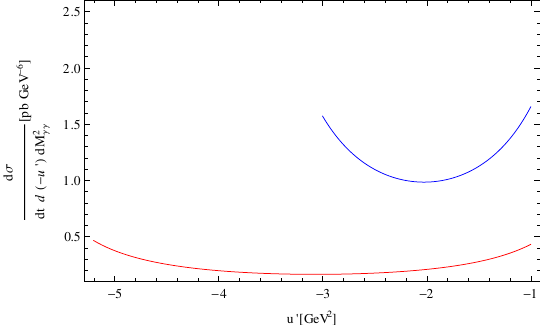}~~~~~~\includegraphics[height=4.5cm]{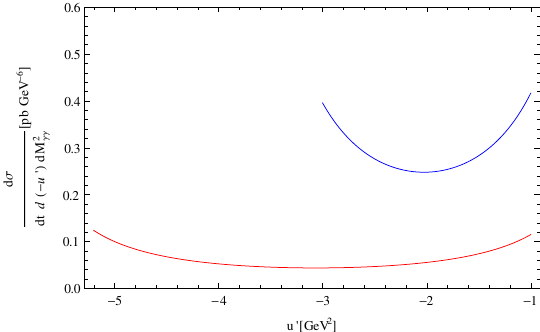}
\caption{The $u'$ dependence of the unpolarized differential cross section $\frac{d\sigma}{dM_{\gamma\gamma}^2du'dt} $ at $t=t_{min}$ and $S_{\gamma N} = 20$ GeV$^2$, for $M_{\gamma\gamma}^2= 4$ GeV$^2$ (blue upper curve) and  for $M_{\gamma\gamma}^2= 6$ GeV$^2$ (red lower curve), for a proton target (left panel) and a neutron target (right panel). }
\label{SigMgg4Mgg6}
\end{figure}
\section{Differential cross section}
We now estimate cross-sections by using a definite model of generalized parton distributions \cite{GK}.
 Although we believe that  the main theoretical uncertainty of our study is the neglected next to  leading order QCD corrections, we may  roughly quantify the model dependence of our result by using another set of GPDs; we illustrate this study on Fig. \ref{VGG_vs_GK} by using  the GPDs of Ref. \cite{VGG}, we get a $20$ per cent decrease (relative to the use of  \cite{GK}) of the cross section at $S_{\gamma N} = 20$ GeV$^2$ and $M_{\gamma\gamma}^2= 4$ GeV$^2$ for all values of $u'$.

Choosing as independent kinematical variables  $\{t,u', M^2_{\gamma\gamma}\}$, the fully unpolarized differential cross section reads
\begin{equation}
\label{crossec}
\frac{d\sigma}{dM^2_{\gamma\gamma}dtd(-u')}=\frac{1}{2}\frac{1}{(2\pi)^3 32 S^2_{\gamma N}M^2_{\gamma\gamma}}\sum_{\lambda,\lambda_1\lambda_2,s_1,s_2}\frac{|\mathcal{T}|^2}{4}
\end{equation}  
 where $\frac{|\mathcal{T}|^2}{2}$ is given in (\ref{tau2}). We show on Fig. \ref{SigMgg4Mgg6}, this fully differential cross section as a function of $u'$ for  $M_{\gamma\gamma}^2= 4$ GeV$^2$ and $6$ GeV$^2$. The bounds in $u'$ are chosen so that both $-u'$ and $-t'$ are larger than $1$ GeV$^2$, leading to the range $1<-u'< M^2_{\gamma \gamma}+1-t$.
 
\begin{figure}[h!]
\centering
\includegraphics[height=8cm]{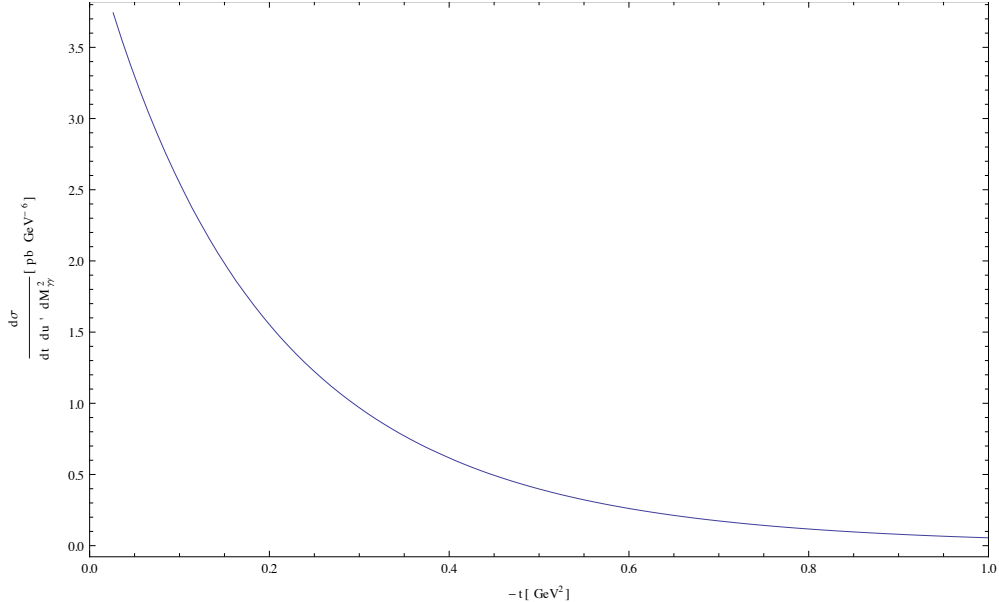}
\caption{The $t-$dependence of the unpolarized differential cross section $\frac{d\sigma}{dM_{\gamma\gamma}^2du'dt} $ at $u' = -2 $GeV$^2$ and $S_{\gamma N} = 20$ GeV$^2$, for $M_{\gamma\gamma}^2= 3$GeV$^2$. }
\label{cs_t}
\end{figure}

The $t-$dependence of the cross section is interesting with respect to the transverse tomography of the nucleon \cite{impact}. We show on Fig. \ref{cs_t} the differential cross-section on a proton as a function of $t-$ for a particular but representative choice of kinematics, namely $M_{\gamma\gamma}^2= 3$GeV$^2$, $u' = -2 $GeV$^2$ and $S_{\gamma N} = 20$ GeV$^2$.

\begin{figure}[h!]
\centering
\includegraphics[height=5cm]{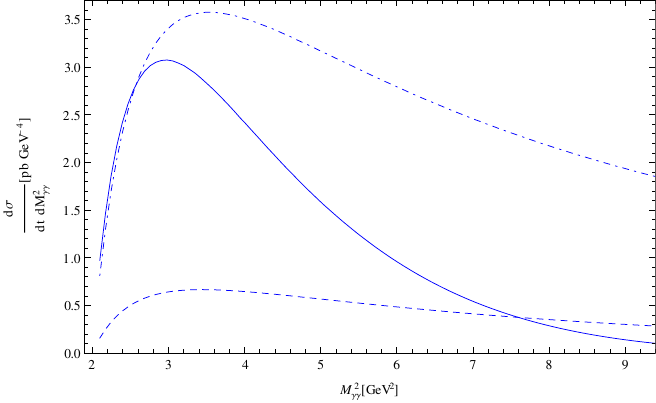} ~~~\includegraphics[height=5cm]{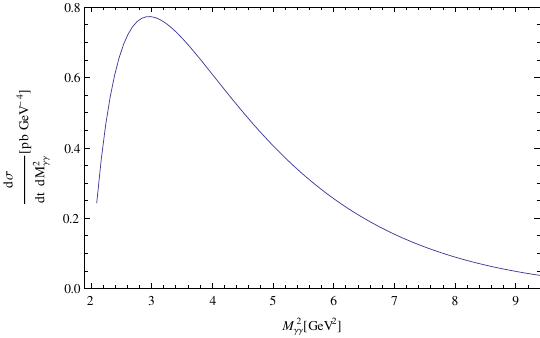}
\caption{The $M_{\gamma\gamma}^2$ dependence of the unpolarized differential cross section $\frac{d\sigma}{dM_{\gamma\gamma}^2dt} $ on a proton(left panel) and on a neutron(right panel) at $t=t_{min}$ and $S_{\gamma N} = 20$ GeV$^2$ (full curves), $S_{\gamma N} = 100$ GeV$^2$  (dashed curve) and $S_{\gamma N} = 10^6$ GeV$^2$  (dash-dotted curve, multiplied by $10^5$). The range of integration with respect to $u'$ is explained in the text.}
\label{cs_M2}
\end{figure}

To get estimates for counting rates let us integrate over $u'$ in this range, as
\begin{equation}
\left.\int_{-u'_{min}}^{-u'_{max}}\frac{d\sigma}{dM_{\gamma\gamma}^2dt(-du')}\right|_{t=t_{min}(M_{\gamma\gamma}^2)}d(-u')=\left.\frac{d\sigma}{dM_{\gamma\gamma}^2dt}\right|_{t=t_{min}(M_{\gamma\gamma}^2)}.
\end{equation}
We show on Fig.\ref{cs_M2} this differential cross section as a function of $M_{\gamma\gamma}^2$ at $\Delta_T =0$ and $S_{\gamma N} = 20$ GeV$^2$, $100$ GeV$^2$ and $10^6$ GeV$^2$ for the photoproduction on a proton and for $S_{\gamma N} = 20$ GeV$^2$ on a neutron target. The energy dependence is moderate between JLab and COMPASS energy ranges. If we consider  higher energies, we find that the cross section $\frac{d\sigma}{dt dM_{\gamma \gamma}^2}$ decreases roughly as $1/S_{\gamma N}$. We understand this fact by remarking that  our process does not benefit from the growth of gluon GPDs as other processes \cite{PSWUPC} do. The process is  unobservable at very high energies such as those discussed in the  LHeC proposal \cite{AbelleiraFernandez:2012cc}
 with a backscattered photon beam of $0(50$ GeV$)$ colliding on a $7$ TeV proton beam. 

\begin{figure}[h!]
\centering
\includegraphics[height=5cm]{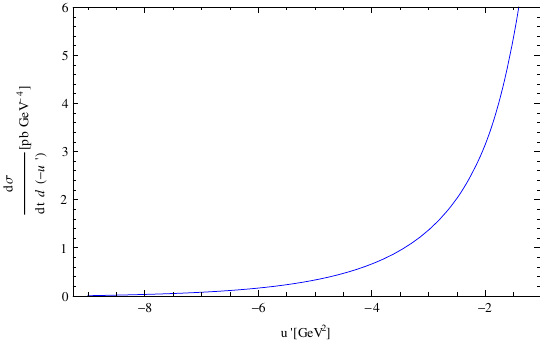} ~~~\includegraphics[height=5cm]{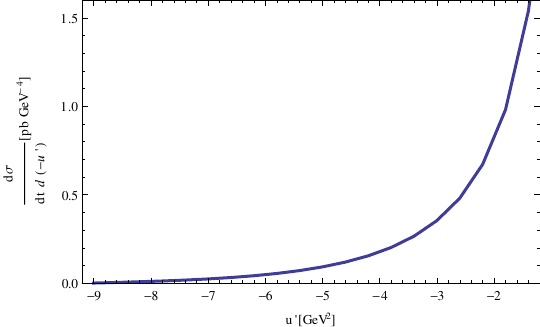}
\caption{The $u'$ dependence of the unpolarized differential cross section $\frac{d\sigma}{dt du'} $ on a proton (left panel) and on a neutron (right panel) at $t=t_{min}$ and $S_{\gamma N} = 20 GeV^2$, integrated over $M_{\gamma\gamma}^2$ as explained in the text. }
\label{csu}
\end{figure}

Integration over $M^2_{\gamma\gamma}$ over the range  $M_{\gamma\gamma}^2 > 2$ GeV$^2$ leads to the $u'$ dependence: 
\begin{equation}
\left.\int_{2} dM_{\gamma\gamma}^2 ~\frac{d\sigma}{dM_{\gamma\gamma}^2dt(-du')}\right|_{t=t_{min}(M_{\gamma\gamma}^2)}=\left.\frac{d\sigma}{dtd(-u')}\right|_{t=t_{min}}
\end{equation}
We show this differential cross section on Fig. \ref{csu} for a proton and neutron target.

The conclusion of these cross-section estimates is straightforward. This reaction can be studied at intense photon beam facilities in JLab. The rates are not very large but of comparable order of magnitude as those for the timelike Compton scattering reaction, the feasibility of which has been demonstrated \cite{Boer}. Since there are no contribution from gluons and sea-quarks, one does not get larger cross-sections at higher energies. Contrarily to timelike Compton scattering \cite{PSWUPC}, it thus does not seem attractive to look for this reaction  in ultra peripheral reactions at hadron colliders. 
\section{Polarization asymmetries}

\subsection{Circular initial photon polarization}
Defining as usual the two circular polarization states of the incoming photon as
\begin{eqnarray}
\epsilon_\bot^\mu(q,+) = - \frac{1}{\sqrt{2}}\left(   \epsilon_\bot^\mu(q,x) + i \epsilon_\bot^\mu(q,y) \right)\,, \;\;\;\;\;\;\epsilon_\bot^\mu(q,-) =  \frac{1}{\sqrt{2}}\left(   \epsilon_\bot^\mu(q,x) - i \epsilon_\bot^\mu(q,y) \right)\,,
\end{eqnarray}
the electromagnetic tensor entering the polarized cross section difference reads
\begin{eqnarray}
\epsilon_\bot^\mu(q,+)\epsilon_\bot^{\nu *}(q,+) - \epsilon_\bot^\mu(q,-)\epsilon_\bot^{\nu *}(q,-)= i\left(  \epsilon_\bot^\mu(q,y) \epsilon_\bot^\nu(q,x) -   \epsilon_\bot^\mu(q,x) \epsilon_\bot^\nu(q,y)   \right) = -i \frac{2}{s}\epsilon^{\mu \nu p n}\,,
\end{eqnarray}
which leads to a cross-section difference
\begin{eqnarray}
{\cal T}_+ {\cal T}^*_+ - {\cal T}_- {\cal T}^*_-    \sim   |\Delta_t|  | p_t|    \sin(\phi)\,.
\end{eqnarray}
The circular polarization asymmetry is thus of $O(\frac{\Delta_T}{Q})$. It is thus inconsistent to calculate it from our formulae since twist 3 contributions that we did not consider, in particular those mandated by electromagnetic gauge invariance \cite{twist3} are of the same order.
\begin{figure}[h!]
\centering
\includegraphics[height=8cm]{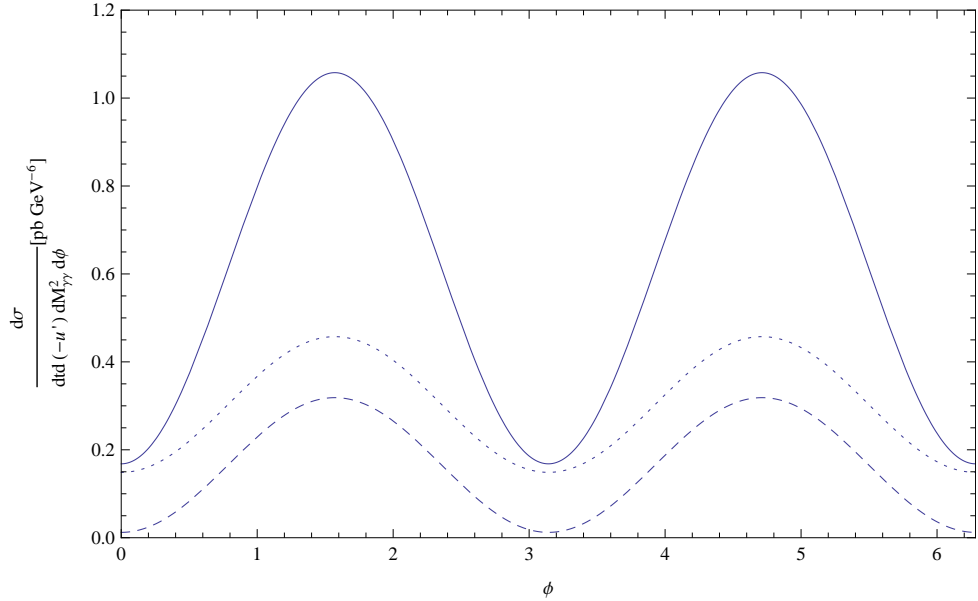}
\caption{The azimuthal dependence of the  differential cross section $\frac{d\sigma}{dM_{\gamma\gamma}^2dt du' d\phi} $ at $t=t_{min}$ and $S_{\gamma N} = 20$ GeV$^2$.  $(M_{\gamma\gamma}^2,u' ) =( 3, -2)$ GeV$^2$  (solid line),  $(M_{\gamma\gamma}^2,u' ) =( 4, -1)$ GeV$^2$  (dotted line) and $(M_{\gamma\gamma}^2,u' ) =( 4, -2)$ GeV$^2$  (dashed line). $\phi$ is the angle between the initial photon polarization and one of the final photon momentum in the transverse plane.}
\label{cs_phi}
\end{figure}

\subsection{Linear initial photon polarization}

Linearly polarized real photons open the way to large asymmetries, as they do for dilepton photoproduction \cite{Goritschnig:2014eba}. Let us consider the case where the initial photon is polarized along the $x$ axis, : $\epsilon(q) = (0,1,0,0)$ and define the azimuthal angle $\phi$ through 
$$
p_T^\mu = ( 0, ~p_T~ cos \phi, ~p_T ~sin\phi, 0).
$$
The cross section exhibits then  an azimuthal  dependence, and one should calculate 
\begin{equation}
\label{crossecLC}
\frac{d\sigma_{l}}{dM^2_{\gamma\gamma}dtd(-u')d\phi}=\frac{1}{2}\frac{1}{(2\pi)^4 32 S^2_{\gamma N}M^2_{\gamma\gamma}}
\sum_{\lambda_1\lambda_2,s_1,s_2}\frac{|\mathcal{T}|^2}{2}.
\end{equation} 
This is shown on Fig. \ref{cs_phi} for different values of $(M_{\gamma\gamma}^2, u' )$ at $ t=t_{min}$ and $S_{\gamma N} = 20$ GeV$^2$. As straightforwardly anticipated, the cross section shows a modulation of the form $A + B ~cos 2\phi$. It turns out that $B$ is negative leading to  a minimum at $\phi=0$ and a maximum at  $\phi=\pi/2$. In all cases, the linear polarization effects are huge. 

\section{Conclusion}
Our calculation of the leading order leading twist amplitude of reaction (\ref{process})  has demonstrated that the photoproduction of a large invariant mass diphoton is an interesting process to analyze in the collinear factorization framework. The amplitude has very specific properties which should be very useful for future GPDs extractions programs e.g. \cite{PARTONS}.

The $O(\alpha_s)$ corrections to the amplitude need to be calculated.They are particularly interesting since they open the way to a perturbative proof of factorization. Moreover, let us recall the reader of the importance of the timelike vs spacelike nature of the probe with respect to the size of the NLO corrections \cite{MPSW}; since the hard scales at work in our process are both the timelike one $M^2_{\gamma\gamma}$ and the spacelike one $ u'$, we are facing an intermediate case between timelike Compton scattering and spacelike DVCS.

In this paper, we only addressed the leading twist amplitude. There are more than one source of $O(\frac{1}{Q})$ corrections to the amplitude. The inclusion of finite hadron mass and finite $t$ effects has been discussed in great detail in \cite{BraunManashov} for the DVCS case and we believe that they can be included in the same way for the reaction studied here. There is an interesting other piece of higher twist contribution which we may solve right away. This is the part related to the twist 2 distribution amplitude  of the final photons. Indeed, this contribution may be read off the formulae written in Ref. \cite{Boussarie:2016qop} where the process $\gamma N \to \gamma \rho(\varepsilon_T) N'$ was calculated. Indeed the real photon DA \cite{photonDA} has the same chiral-odd structure as the transversely polarized $\rho$ meson, and thus its contribution to the scattering amplitude may be deduced from the $\gamma \rho^0_T$ meson case, by replacing $\frac{f_\rho}{\sqrt 2}$ in the normalization of  the meson DA by the constant  $e_q \chi <\bar q q>$ including the magnetic susceptibility $ \chi$ of the QCD vacuum. Theoretical estimates  \cite{photonDA} of $\chi <\bar q q>$ are around $50$ MeV. Using numerical estimates of Ref.  \cite{Boussarie:2016qop}, this leads to a smaller than $10 \%$ correction to the cross section already at $M^2_{\gamma\gamma}=3 $ GeV$^2$. Including the contributions of higher twist GPDs is a tougher job which will need long further studies  \cite{Anikin:2009hk} ; indeed factorization may even break down.

We only addressed the case of a real photon beam. The analysis of the corresponding reaction with quasi real photons obtained from a lepton beam goes along the same lines but needs to be complemented by the analysis of the Bethe Heitler processes where one or two photons are emitted from the lepton line. The relative magnitude of this background to the QCD process is likely to depend much on the energy and on the azimuth of the produced photons, as it is the case in the known example of DVCS. We can anticipate  that the contribution of the double Bethe Heitler process $e^- \to  e^-   \gamma^*(t) ~\gamma~ \gamma$  to the reaction $e^- N \to  e^-    \gamma ~\gamma ~N'$ will dominate at small $t$. This should lead to interesting interference effects with the amplitude that we studied here, as in the well-know case of DVCS. We shall study this case in details in a future work.

\paragraph*{Acknowledgements.}
\noindent
We acknowledge useful conversations with Renaud Boussarie, Michel Guidal, Herv\'e Moutarde and Samuel Wallon.  This work is partly supported by Grant No. 2015/17/B/ST2/01838 from the National Science Center in Poland, by the Polish-French collaboration agreements  Polonium and COPIN-IN2P3, and  by the French Grant ANR PARTONS (Grant No. ANR-12-MONU-0008-01).

\end{document}